\documentclass[10pt,a4paper,final]{iopart}

\usepackage[dvips]{graphicx}
\usepackage{color}
\usepackage{ulem}

\usepackage{dcolumn}

\usepackage{bm}
\usepackage{amssymb}
\def\be{\begin{equation}}
\def\ee{\end{equation}}
\def\bea{\begin{eqnarray}}
\def\eea{\end{eqnarray}}

\def\be{\begin{equation}}
\def\ee{\end{equation}}

\begin{document}

\title[Spin relaxation and the Kondo effect in transition metal dichalcogenide monolayers]{Spin relaxation and the Kondo effect in transition metal dichalcogenide monolayers}

\author{Habib Rostami}
\address{Istituto Italiano di Tecnologia, Graphene Labs, Via Morego 30, I-16163 Genova, Italy}
\address{School of Physics, Institute for Research in Fundamental Sciences (IPM), Tehran 19395-5531, Iran}

\author{Ali G. Moghaddam}
\address{Department of Physics, Institute for Advanced Studies in Basic
Sciences (IASBS), Zanjan 45137-66731, Iran}
\ead{agorbanz@iasbs.ac.ir}

\author{Reza Asgari}
\address{School of Physics, Institute for Research in Fundamental Sciences (IPM), Tehran 19395-5531, Iran}
\address{School of Nano Science, Institute for Research in Fundamental Sciences (IPM), Tehran 19395-5531, Iran}

\begin{abstract}
We investigate the spin relaxation and Kondo resistivity caused by magnetic impurities in doped transition metal dichalcogenides monolayers. We show that momentum and spin relaxation times due to the exchange interaction by magnetic impurities, are much longer when the Fermi level is inside the spin split region of the valence band. In contrast to the spin relaxation, we find that the dependence of Kondo temperature $T_K$ on the doping is not strongly affected by the spin-orbit induced splitting, although only one of the spin species are present at each valley. This result, which is obtained using both perturbation theory and poor man's scaling methods, originates from the intervalley spin-flip scattering in the spin-split region. We further demonstrate the decline in the conductivity with temperatures close to $T_K$ which can vary with the doping. Our findings reveal the qualitative difference with the Kondo physics in conventional metallic systems and other Dirac materials.
\end{abstract}

% Uncomment for PACS numbers
\maketitle

%%%%%%%%%%%%%%%%%%%%%%%%%%%%%%%%%%%%%%%%%%%%%%%%%%%%%%%%%%%%%%%%%%%%%%%%%%%%%%%%

\section{introduction}\label{sect:intro}
Isolation of two dimensional (2D) materials during the last decade seems to revolutionize the common understandings in condensed matter physics \cite{novo05pnas,xu13,geim13,butler13}. Graphene as the most known of these materials, besides attracting much interest from a fundamental point of view, has opened a new route to application in electronic devices \cite{geim07,neto09}. On the other hand, despite lacking intrinsic magnetic properties due to the long spin relaxation time, graphene has been suggested to be used in spintronics \cite{tombros07,burkard07,fabian14}. Very soon after graphene, the transition-metal dichalcogenides (TMDs) with layered van der Waals structure, have been exfoliated to few- and single-layer crystalline samples \cite{mak10,wang10,chhowalla}. They exhibit diverse electronic characteristics starting from metallic and semiconducting to more exotic phases, including superconducting and even charge density wave \cite{wang12}.
As a general characteristic, most of theses materials reveal a very large intrinsic spin-orbit interaction (SOI) which has made them very promising for spintronic applications \cite{zhu11,yuan13,riley14,xu14,burkard14prx,roldan14-2d,loss13}. Intriguingly, the combination of the large SOI with two-valley band structure and the presence of a direct band gap, make a monolayer of TMDs as a building block for valleytronics and optoelectronics. In particular, using circularly polarized light, various groups observed an optically induced valley polarization \cite{zeng-mak,cao12}, which eventually enabled the experimental verification of the valley Hall effect \cite{mceuen14,niu07,xiao12}.
\par
Various groups have studied spin dependent effects in TMDs as well as their magnetic properties not only because of possible application, but also due to the fundamental interests. It has been predicted that Ruderman-Kittel-Kasuya-Yosida interaction between magnetic impurities consists of Heisenberg, Ising and Dzyaloshinsky-Moriya terms \cite{Parhizgar13} owing to the large spin splitting of the valence band and hexagonal symmetry of monolayer MoS$_2$. It has been found, on the other hand, that the spin-orbit mediated spin relaxation results in spin lifetimes larger than nanoseconds \cite{roldan13}. These relaxation mechanisms are consequences of the Rashba spin-orbit coupling and Dyakonov-Perel contribution induced by interactions and impurity scattering \cite{WW14}. Other studies have indicated that only the intervalley electron-phonon scattering has a significant effect on the intrinsic in-plane spin relaxation among other scattering mechanism \cite{song13}.
Besides spin relaxation phenomena caused by the SOI, different groups have explored the emergence of magnetic phases in TMDs. In particular, {\it ab initio} calculations for MoS$_2$ have suggested that doping with either transition metals including Mn, Fe, Co, Re or vacancies and defects could make the system ferromagnetic \cite{cheng13,vaveh13,yakobson,idrobo,
yan14,Sachs13}. Moreover, theoretical and experimental studies have found signatures of magnetism at the edges and grain boundaries of these materials \cite{sullivan,chen08,botello,tongay12}.
\par
One of the key properties of TMDs monolayers (ML) is the large spin splitting in the valence band which originates from the SOI and changes its sign between two inequivalent valleys \cite{xiao12}. This property would affect the spin dependent scattering phenomena, for instance, in the presence of magnetic impurities. So motivated by this fact, we will address the magnetic impurity problem in doped TMDs-ML with particular focus on the low temperatures and at the vicinity of the Kondo regime. \emph{The Kondo effect}, in its orthodox form, is revealed as a logarithmic upturn in the low temperature resistivity of metals dilutely doped with magnetic impurities \cite{kondo}. This behavior originates from the strong antiferromagnetic correlation between conduction electrons and the localized magnetic moments which are coupled to each other with a constant ${\cal J}$ \cite{hewson}. As a first attempt, Jun Kondo employed the perturbation theory to obtain the $\log T$ contribution in the resistivity. His theory was able to explain the appearance of the resistance minimum at a temperature $T_{\rm min}$ due to the competition between the phonon and nonmagnetic impurities contributions to that of magnetic impurities. By extensions of the perturbation approach, Abrikosov summed up the leading logarithmic contributions from the higher order scattering processes which led to a divergent behavior for the resistivity at the so-called Kondo temperature $T_K$ \cite{abrikosov}. This observation was a signal indicating strong scatterings of the conduction electrons by the local magnetic moments. From the breakdown point of the perturbation, a non-analytical dependence of $T_K$ on the coupling constant ${\cal J}$ with the form $T_K={\cal D}\exp(-1/|\rho {\cal J}|)$ was found, in which $\rho$ is the density of states (DOS) of the conduction electrons. Using the more powerful methods which were developed afterwards, the exponential dependence of the Kondo temperature on the inverse of coupling constant was verified and precise form of the prefactor ${\cal D}$ was obtained. On the other hand, more satisfactory solutions of \emph{the Kondo problem} revealed that the local moments are gradually screened by the conduction electrons at very low temperatures $T<T_K$. This picture has been confirmed during 70's and 80's by a variety of techniques in particular the renormalization group (RG) and exact methods among others \cite{anderson,wilson,andrei,nozieres,tsvelick}.
\par
In this paper, by exploiting the perturbation theory and poor man's renormalization method of Anderson and focusing on the regime of temperatures above Kondo ($T\gtrsim T_K$), the onset of the Kondo effect in TMDs-ML is studied. We find that the low-order spin dependent scatterings are suppressed in the spin-split region of the band structure while they remain strong at higher doping strengths when the Fermi level crosses both spin subbands. In contrast, the Kondo temperature is not strongly affected even when the Fermi level is inside the spin split region and crosses only one spin subband at each valley. This result is also verified using the poor man's scaling analysis which is very similar in spirit to the Wilson's numerical RG method \cite{anderson,wilson}. To understand these results we notice that in the vicinity of $T_K$ there are equal probabilities for inter- and intra-valley scatterings owing to the short-range exchange interactions with magnetic impurities. Subsequently the Kondo characteristics in the case in which the Fermi level lies inside the spin split region are more or less the same as heavily doped cases in which the Fermi level is deep inside the valence or conduction band. Our findings indicate some differences with the investigations of the Kondo effect in other 2D systems such as graphene \cite{uchoa11,vojta,fuhrer,fuhrer-cmt} as well as those in the presence of the SOI \cite{zarea12,sandler16,vojta13,isaev15}. Besides fundamental importances, these results could be useful for the future investigations about the spintronic applications of magnetically doped TMDs-ML.
\par
This paper is organized as follows. In Sec. \ref{sec2}, the TMDs-ML Hamiltonian is given and the Kondo model is obtained from the Anderson model via Schrieffer-Wolff transformation. To be precise, we will focus on monolayer molybdenum disulfide (ML-MoS$2$) as a prototype of these systems. Then the Boltzmann transport method for the scattering rate calculations and poor man's scaling technique are are presented. In Sec. \ref{sec3} the numerical results for the spin life time owing to the scattering from magnetic impurities as well as a Kondo temperature variation with the chemical potential and the impurity levels are obtained.
In addition the magnetic impurities contribution in the conductivity is obtained. This part is closed by discussion on the results and their relation to those in other 2D materials. Finally, Sec. \ref{sec4} is devoted to the conclusions.

\section{Theoretical model and formalism}\label{sec2}
We start by writing down the low-energy band structure of ML-MoS$_2$ which consists of two valleys, $K$ and $K'$, which are time-reversal with respect to each other. In the vicinity of each valley, we have an almost spin-degenerate conduction band and two spin-split valence subbands separated by the spin-orbit coupling $2\lambda \sim 160 $ meV. Having known from first principle calculations, the conduction band is mainly formed from Molybdenum $d_{z^2}$ orbitals while the valence band is constructed by the $\{d_{x^2-y^2},d_{xy}\}$ orbitals of Mo with mixing from $\{p_x,p_y\}$ of S.  Verified by both tight-binding models and ${\bf k}\cdot {\bf p}$ method and symmetry arguments, the low-energy excitation energy of the ML-MoS$_2$ is provided by
the modified Dirac Hamiltonian around valley labeled by $\tau=\pm1$ and for spin $s=\pm1$ \cite{rostami13,kormanyos13,liu13,kormanyos15},
\begin{eqnarray}
\nonumber{\cal H}_{\tau s}({\bf k})=\frac{\Delta}{2}\sigma_z+\lambda\tau s
\frac{1-\sigma_z}{2}
\\
+
t_0 a_0 {\bf k}\cdot{\bm \sigma}_\tau +\frac{\hbar^2|{\bf k}|^2}{4m_0}(\alpha+\beta\sigma_z)
\end{eqnarray}
Here, the Pauli matrices ${\bm \sigma}_\tau=(\tau\sigma_x,\sigma_y)$ and $\sigma_z$ operate in a space defined by conduction and valence bands.

The eigenvalues of this Hamiltonian are given by
\begin{eqnarray}\label{dispersion}
\epsilon_{k\tau s}&=&\frac{1}{2}\lambda\tau s+\frac{\hbar^2 k^2}{2m_0}\alpha \nonumber\\ &\pm&\sqrt{(\frac{\Delta-\lambda\tau s}{2}+\frac{\hbar^2k^2}{2m_0}\beta)^2+(t_0a_0)^2k^2}.
\end{eqnarray}
As it will be clear in the following, the Kondo temperature and related transport properties are mostly influenced by the density of states corresponding to the low-energy bands. Then we use the following approximate form of the DOS for each spin index,
\begin{equation}
\label{dos}
\rho_{s} (\epsilon) \approx |\epsilon| \sum_{\tau }
\Big \{
\rho^{\rm c}_0  \Theta \Big (\epsilon-\frac{\Delta} {2} \Big )+ \rho^{\rm v}_0  \Theta \Big (-\epsilon-\frac{\Delta}{2}+\lambda \tau s  \Big )
\Big \}
\end{equation}
with constants $\rho^{\rm c}_0 \approx 0.016~({\rm eV} a_0 )^{-2}$  and $\rho^{\rm v}_0 \approx 0.029~({\rm eV} a_0)^{-2}$ which are different basically due to the difference in the electron and hole effective masses \cite{footnote}. The other important band parameters are the semiconducting gap $\Delta= 1.9$ eV and the intrinsic SOI strength $\lambda=80$ meV.
\par

In order to study the effect of dilute magnetic impurities, we begin with the Anderson Hamiltonian as,
\begin{eqnarray}
{\cal H}&=&{\cal H_{\rm bath}}+{\cal H_{\rm imp}}+{\cal H}_V , \nonumber\\
{\cal H_{\rm bath}}&=&\sum_{k\tau s}{\epsilon_{k \tau s}\hat n_{k \tau s}} ,\nonumber\\
{\cal H_{\rm imp}}&=&\sum_s{\epsilon_d \hat n_{ds}+U\hat n_{d\uparrow}\hat n_{d\downarrow}},\nonumber\\
{\cal H}_V&=&\sum_{k\tau s}{\{V_{k \tau d} c^\dagger_{k\tau s}c_{ds}+V^\ast_{k\tau d}c^\dagger_{ds}c_{k\tau s}}\},
\end{eqnarray}
where $U$ is the on-site Coulomb repulsion and $V_{k \tau d}$ is spin-conserving hybridization between impurity level and the charge carriers in ML-MoS$_2$. The impurity level is assumed to have energy $\epsilon_d$ and the corresponding number operator is denoted by $\hat n_{ds}$ for the spin $s$ impurity state. The energy dispersion $\epsilon_{k \tau s}$ of the host electrons in doped ML-MoS$_2$ is given by Eq. (\ref{dispersion}). All energies, here, are measured from the middle of the energy gap. Since we are dealing with the low-energy model of TMDs-ML, it is consistent to consider the hybridization between impurity and the itinerant electrons in the Dirac delta function form $V\delta(r-R)$. Subsequently, there will be no momentum dependence in $V_{k\tau d}$ in the orbital basis model Hamiltonian of the system. Although the projection of the band basis model (e.g. the two-band model) might result in a momentum dependence in the hybridization, we assume a constant value for it as a reasonable approximation due to the considerably larger energy band gap of MoS$_2$. In addition, the hybridization in the lattice models when the impurity is exactly placed in a transition metal site should be constant in analogy with the case of graphene when impurities on top of carbon atoms \cite{uchoa11}. One should note that owing to the similar electronic structure, the magnetic impurities are energetically preferred to replace a transition metal atom of TMD.
\par
In order to describe the local-moment exchange interactions with the conduction electrons, we utilize the Schrieffer-Wolff transformation, when only the spin doublet state of the impurity is kept, which results in an effective Hamiltonian $\tilde{\cal H}={\cal H_{\rm bath}}+{\cal H_{\rm imp}}+\tilde{\cal H}_2$ in which the last term contains all second order terms in $V_{k \tau d}$,
\begin{eqnarray}
\tilde{\cal H}_2={\cal H}'_{\rm imp}+{\cal H}_{\rm ch}+{\cal H}_{\rm dir}+{\cal H}_{\rm ex}.
\end{eqnarray}
The four terms correspond to corrections to the impurity Hamiltonian, charge fluctuations, spin independent impurity-bath coupling, and exchange interaction between impurity and bath electrons, respectively. The detailed form of the correction terms of the systems like ML-MoS$_2$ with the spin and valley degree of freedom can be found in \ref{app-a}.
Assuming single occupation regime of the impurity state ($\hat n_{ds}+\hat n_{d\bar s}=1$), the only leading term will be the so-called $s-d$ exchange or Kondo Hamiltonian,
\begin{eqnarray}\label{exchange}
{\cal H}_{\rm ex}=&&-\frac{1}{2}\sum_{k,k',\tau,\tau'}\{
[I^{\tau'\tau}_{k'k\uparrow}c^\dagger_{k'\tau'\uparrow}c_{k\tau\uparrow}-I^{\tau'\tau}_{k'k\downarrow}c^\dagger_{k'\tau'\downarrow}c_{k\tau\downarrow} ] S_{z}\nonumber \\
&&+
J^{\tau'\tau}_{k'k\uparrow}c^\dagger_{k'\tau'\uparrow}c_{k\tau\downarrow} S_{-}+
J^{\tau'\tau}_{k'k\downarrow}c^\dagger_{k'\tau'\downarrow}c_{k\tau\uparrow}S_{+}\},~~~~~~
\end{eqnarray}
with $S_z=(n_{d\uparrow}-n_{d\downarrow})/2$, $S_{+}=c^\dagger_{d\uparrow}c_{d\downarrow}$, and $S_{-}=c^\dagger_{d\downarrow}c_{d\uparrow}$ defining the local spin operators of impurity.
\subsection{The lifetimes and T-matrix method}
In order to address the Kondo effect and subsequently spin lifetime owing to the magnetic scattering, we use the Boltzmann transport theory to obtain the contribution of magnetic impurities in the resistivity. Working on the scheme of relaxation time approximation, the key quantity appearing in the
Drude conductivity formula is the transport lifetime given by,
\begin{eqnarray}\label{tau}
\frac{1}{ \tau_p (\epsilon)}&=&n_{\rm imp}\frac{2\pi}{\hbar}\sum'_{\tau s}\sum_{\tau's'}\int{\frac{d^2 k'}{(2\pi)^2}|\langle k' \tau's'|T|k\tau s \rangle|^2}
 \nonumber\\ &\times& \Big [1- \cos\left (\theta_{{\bf k}' {\bf k}}\right ) \frac{k'}{k} \Big ]
\delta(\epsilon-\epsilon_{k'\tau's'}),
\end{eqnarray}
where $\sum'$ sums over the spin and valley index, which satisfies $\epsilon_{k\tau s}=\epsilon$, $\theta_{{\bf k}' {\bf k}}$ is the angle between ${\bf k}'$ and ${\bf k}$ vectors and $n_{imp}$ indicates the density of magnetic impurities. In particular, for the energies lie within the spin-split region ($-\Delta/2-\lambda<\epsilon<-\Delta/2+\lambda$) only the two subbands satisfying $\tau s=1$ are allowed. Otherwise, all four different subbands can contribute in the scattering and transport through the system. In the above relation, $\langle k' \tau's'|T|k\tau s \rangle$ indicates the $T$-matrix element responsible for the scattering amplitude between many-body states in which a particle from the state $|k \tau s \rangle$ is scattered to another state $|k' \tau' s' \rangle$.
\begin{figure}[tp]
\includegraphics[width=0.85\linewidth]{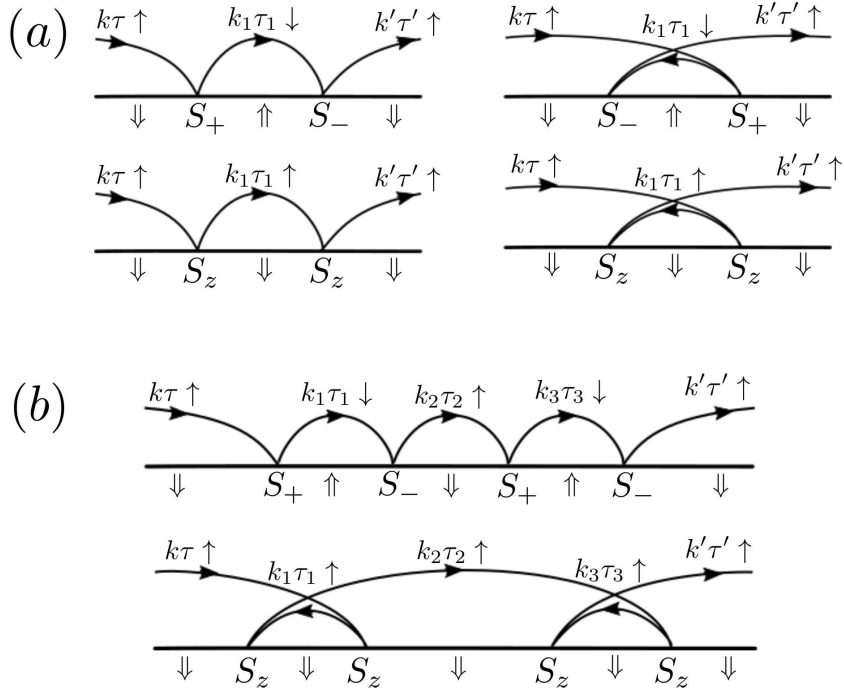}
\caption{(a) Second order diagrams corresponding to spin conserving scatterings $T_{\uparrow\uparrow}^{(2)}$ due to the exchange interaction with impurity spin. (b) Ladder-like forth order diagrams which contribute in $T_{\uparrow\uparrow}^{(4)}$. Assuming all higher order diagrams of this kind and summing over their contribution leads to the parquet approximation for the scattering rates. Similar diagrams can be considered for the other three scattering terms $T_{\downarrow \downarrow}$, $T_{\downarrow \uparrow}$, and $T_{\uparrow \downarrow}$.}
\label{fig1}
\end{figure}
\par
We use the perturbative Dyson series to obtain the $T$-matrix elements as below,
\begin{eqnarray}
&&\langle k' \tau' s'|T|k\tau s\rangle=\langle k' \tau' s'|{\cal H}_{\rm ex}|k\tau s\rangle \nonumber\\
&&+\langle k' \tau' s'|{\cal H}_{\rm ex} G_0^+(\epsilon_{k\tau s}) {\cal H}_{\rm ex}|k\tau s\rangle +\cdots
\end{eqnarray}
where an unperturbed retarded Green's function is given by $G_0^+(\epsilon_{k\tau s})=(\epsilon_{k\tau s} -{\cal H}'_{\rm imp} +i\eta)^{-1}$.
Since the scattering out of magnetic impurity governed by $s-d$ exchange Hamiltonian is isotropic irrespective of momentum directions, it turns out that the $T$-matrix elements depends only on the energy and the spin directions of the incoming and outgoing electronic states. Therefore, we will denote them simply by $T_{s's}(\epsilon)\equiv \langle k' \tau' s'|T|k\tau s\rangle$.
\par
To lowest order in coupling constants ${\cal J}$ and ${\cal I}$, the matrix elements $T_{s's}^{(1)}=\langle k' \tau' s'|{\cal H}_{\rm ex}|k\tau s\rangle$ between different spin directions and independent of electron energy are given by,
\begin{eqnarray}
&&T^{(1)}_{\uparrow\uparrow}  =-\frac{\Omega}{2}I^{\tau'\tau}_{k'k\uparrow}S_z
~~,~~~~T^{(1)}_{\downarrow\downarrow}=\frac{\Omega}{2}I^{\tau'\tau}_{k'k\downarrow}S_z
\nonumber\\
&&T^{(1)}_{\uparrow\downarrow} =-\frac{\Omega}{2}J^{\tau'\tau}_{k'k\uparrow}S_{-} ~,~~~T^{(1)}_{\downarrow\uparrow} =-\frac{\Omega}{2}J^{\tau'\tau}_{k'k\downarrow}S_{+},
\end{eqnarray}
with $\Omega \propto a_0^2$ indicating the unit cell area. As explained in \ref{app-b} working in the validity range of perturbation theory, one can make further simplifications which lead to the replacement of all coefficients $I^{\tau'\tau}_{k'k s}$ and $J^{\tau'\tau}_{k'ks}$ with a single effective coupling constant,
\begin{equation}
{\cal J}(\epsilon,\epsilon_d )=2|V|^2\left[\frac{1}{\epsilon-\epsilon_d -U}-\frac{1}{\epsilon-\epsilon_d }\right].\label{J}
\end{equation}
Subsequently, the second order contributions to the $T$-matrix finally read,
\begin{eqnarray}
T^{(2)}_{\uparrow\uparrow}          &=& \frac{\Omega{\cal J}(\epsilon,\epsilon_d)^2}{4}\int{d\omega} \rho(\omega) \frac{S^2+S_z[2f(\omega)-1]}{\epsilon-\omega+i\eta},
\nonumber\\
%T^{(2)}_{\downarrow\downarrow} &=& \frac{\Omega{\cal J}(\epsilon,\epsilon_d)^2}{4}\int{d\omega} \rho(\omega) \frac{S^2-S_z[2f(\omega)-1]}{\epsilon-\omega+i\eta},
%\nonumber\\
T^{(2)}_{\uparrow\downarrow}      &=& \frac{\Omega{\cal J}(\epsilon,\epsilon_d)^2}{4}\int{d\omega} \rho(\omega) \frac{S_- [2f(\omega)-1]}{\epsilon-\omega+i\eta},
%\nonumber\\
%T^{(2)}_{\downarrow\uparrow}      &=& \frac{\Omega{\cal J}(\epsilon,\epsilon_d)^2}{4}\int{d\omega} \rho(\omega) \frac{S_+[2f(\omega)-1]}{\epsilon-\omega+i\eta}.
\end{eqnarray}
and with silmilar relations for $T_{\downarrow\downarrow}$ and $T_{\downarrow\uparrow}$ in which $S_z$ and $S_-$ are repalced with $-S_z$ and $S_+$, respectively.
\par
Inserting the obtained expression for the first and second order $T$-matrix elements which depend now on the single coefficient ${\cal J}$ and using the relaxation time formula (\ref{tau}), we reach for the
following result for the transport lifetime which depends only on the energy $\epsilon$ as,
\begin{equation}
\frac{1}{\tau_p (\epsilon)}=\frac{n_{\rm imp}}{2\pi \hbar}\rho(\epsilon) \sum'_{\tau s}{ \langle |T_{ss}(\epsilon)|^2 +|T_{\bar ss}(\epsilon)|^2\rangle_{\rm imp} }.
\end{equation}
Then from first and second order terms of the T-matrix, the life-time up to forth order in ${\cal J}$ reads,
\begin{eqnarray}
\frac{1}{\tau_p (\epsilon)} &\approx& \frac{3 g \Omega^2}{2\pi\hbar}  n_{\rm imp}\rho(\epsilon)[\frac{{\cal J}(\epsilon,\epsilon_d)}{4}]^2
\left\{ 1-{\cal J}(\epsilon,\epsilon_d)
\right. \nonumber \\  &\times& \left.
{\rm Re}[2\Gamma_2(\epsilon,T)-\Gamma_1(\epsilon)]
\right\}
+{\cal O}[{\cal J}(\epsilon,\epsilon_d)^4]\label{3rd-order}
\end{eqnarray}
Note that $\bar s=-s$ and $\langle\cdots \rangle_{\rm imp}$ is used to describe the averaging over the direction of randomly oriented impurity spins. The appearance of $g$ here originates from the constraint $\epsilon=\epsilon_{k \tau s}$, over summation, which determines the number of subbands at the energy $\epsilon$. In addition the two integrals $\Gamma_{1,2}$ have the relations,
\begin{eqnarray}
&&\Gamma_{1}(\epsilon)=\int_{-D_0}^{D_0} {d\omega}\rho(\omega)/(\epsilon-\omega+i\eta), \\
&&\Gamma_{2}(\epsilon,T)=\int_{-D_0}^{D_0} {d\omega}f(\omega)\rho(\omega)/(\epsilon-\omega+i\eta),
\end{eqnarray}
\par
The Kondo characteristic in the resistivity originates from many-body higher order processes. So, one needs to go beyond the lowest order the single particle formalism. All of the processes up to any order can be formally described using Feynman diagrams. For instance, Fig. \ref{fig1}(a) shows the four possible diagrams corresponding to the second order terms in which the spin is conserved keeping the conduction electron and the impurity in up and down states, respectively, at the end of process.
Working in the so-called parquet approximation scheme for the Kondo effect \cite{Verwoerd,solyom,Aji}, we only consider ladder-like diagrams.
Then the higher order corrections involving logarithmic temperature-dependent factors are taken into account under this approach.
Fortunately, as one can simply understand from the diagrams of Fig. \ref{fig1}(b) and their higher order extensions, these terms result in a geometric progression.
So, by summation of higher order contributions in Eq. (\ref{3rd-order}) under the parquet approximation, the transport lifetime reads,
\begin{equation}
\frac{1}{\tau_p (\mu)}=\frac{1 }{2\pi\hbar} \frac{3 g   n_{\rm imp}\rho(\epsilon)\Omega^2 [{\cal J}_0/4]^2}{1+{\cal J}_0{\rm Re}[2\Gamma_2(\mu,T)-\Gamma_1(\mu)] },\label{parquet}
\end{equation}
in which the notation ${\cal J}_0={\cal J}(\mu,\epsilon_d)$ is used.
\par
The Kondo temperature $T_K$ under this scheme can be defined as the temperature in which the spin-flip scattering rate and the resulting contribution in the resistivity become divergent and the perturbation breaks down \cite{solyom}. Therefore, $T_K$ can be extracted from the criterion, $1+{\cal J}_0{\rm Re}[2\Gamma_2(\mu,T_{K})-\Gamma_1(\mu)]=0$. In general, the Kondo temperature is only a function of the density of states of the MoS$_{2}$ and the exchange interaction with the system.  We should remind that $\Gamma_2$ depends explicitly on temperature. Since both $\Gamma_{1,2}$ are functions of the chemical potential as well, we expect $T_{K}$ to be a function of the chemical potential ${\mu}$ and the impurity state single occupation energy $\epsilon_d$. In the following section, we will use the aforementioned criteria to extract the Kondo temperature and its variations with both $\mu$ and $\epsilon_d$.
\subsection{Spin relaxation time}
Another quantity which can be considered here is the spin relaxation caused
by the magnetic impurities. In the Boltzmann scheme and taking into account only spin flipping processes, the spin relaxation time $\tau_s$ is given by,
\begin{eqnarray}
\frac{1}{\tau_s}=\frac{\sum'_{k\tau}\sum_{k'\tau'}{\frac{\partial f(\epsilon_{k\tau \downarrow})}{\partial\mu}W_{k'\tau'\uparrow,k\tau\downarrow}}}{\sum_{k\tau}\frac{\partial f(\epsilon_{k\tau \downarrow})}{\partial\mu}},
\end{eqnarray}
where spin-flip scattering rate is given by
\begin{eqnarray}
W_{k'\tau'\uparrow,k\tau\downarrow}=\frac{ g n_{\rm imp}}{2\pi \hbar}\delta(\epsilon_{k\tau \downarrow}-\epsilon_{k'\tau' \uparrow})| T_{\uparrow\downarrow}|^2.
\end{eqnarray}
Implementing some straightforward algebra, it can be shown that the spin relaxation time is indeed proportional to the transport lifetime (or the momentum relaxation time) given by,
\begin{eqnarray}
\frac{1}{\tau_s}=\frac{g }{2\pi\hbar}n_{\rm imp} \rho(\varepsilon)|T_{\uparrow\downarrow}|^2=\frac{2}{3}\frac{1}{\tau_p}.
\end{eqnarray}
The appearance of numerical prefactor $2/3$ can be interpreted as the fact that scattering rate of spin-flipping processes are $2/3$ of the whole scattering rate.
\subsection{Poor man's scaling analysis}
As an alternative method one can use poor man's scaling analysis introduced by Anderson which leads to the renormalization equation of the coupling constant ${\cal J}$ due to the exchange coupling to the electron bath \cite{anderson}. In this part, the electron bath is modeled with a single fictitious band extended over $(-D+\mu,D+\mu)$. The choice of symmetrically extended band around Fermi level is suitable to obtain the renormalization relations. However, this assumption does not cause any crucial limitation of the method, since the explicit form of the DOS $\rho(\epsilon)$ is taken into account. Following the method of Withoff and Fradkin, the RG flow equations for the coupling constant reads, \cite{fradkin}
\begin{equation}
\frac{d{\cal J}}{d\ln D}= {\cal J} \frac{D}{\tilde{\rho}(D)} \frac{d\tilde{\rho}(D)}{dD} +{\cal J}^2 \tilde{\rho}(D)
\label{rg-flow}
\end{equation}
with $\tilde{\rho}(D)=\rho(D)+\rho(-D)$.
\par
The explicit form of the RG flow equations for TMDs-ML can be obtained invoking the DOS given by Eq. (\ref{dos}). By integrating over the band width $D$ from $D_0$, which is the initial value of the cutoff, to the low-energy cutoff $D_K=k_BT_K$ in which the coupling constant diverges, the Kondo temperature will be obtained. We already know that we can consider three different regimes for the doping of TMDs depending on the position of the chemical potential, namely: $n$-doped, intermediate $p$-doped when only one of the spin split subbands is intersected, and heavily $p$-doped. For the sake of simplicity in the first case, only the contribution of the conduction band is taken into account while in the second and third cases, one of the valence subbands or both are considered. Therefore, we always have a fictitious band over the range $(-D+\mu, D+\mu)$, while only a part of it has finite DOS. So, there will be certain range of cutoffs for which only one edge of the fictitious band has a finite contribution. On the other hand, when the cutoff $D$ is made small enough under RG, both edges of the band will contribute in the Kondo scattering. Hence, the scaled fictitious band lies completely inside either conduction or valence band and subsequently $\tilde{\rho}(D)$ becomes constant for small enough cutoffs $D<|\mu|-\Delta/2$. So at the final stage of the renormalization, the flow equation takes the simple form $d{\cal J}/d\ln D \propto {\cal J}^2$.

\section{Numerical Results and Discussion}\label{sec3}
In this section, we will present our numerical results for the relaxation times, Kondo temperature and the conductivity variation with temperature. Then we will discuss over the obtained results and their relevance for the experiments and promising applications of TMDs-ML in spintronics.
Throughout this article, we use typical values for the Anderson model parameters as $V=0.5$ eV, $D_0=U=10$ eV.
\subsection{Spin relaxation}
\begin{figure}[t]
\includegraphics[width=0.8\linewidth]{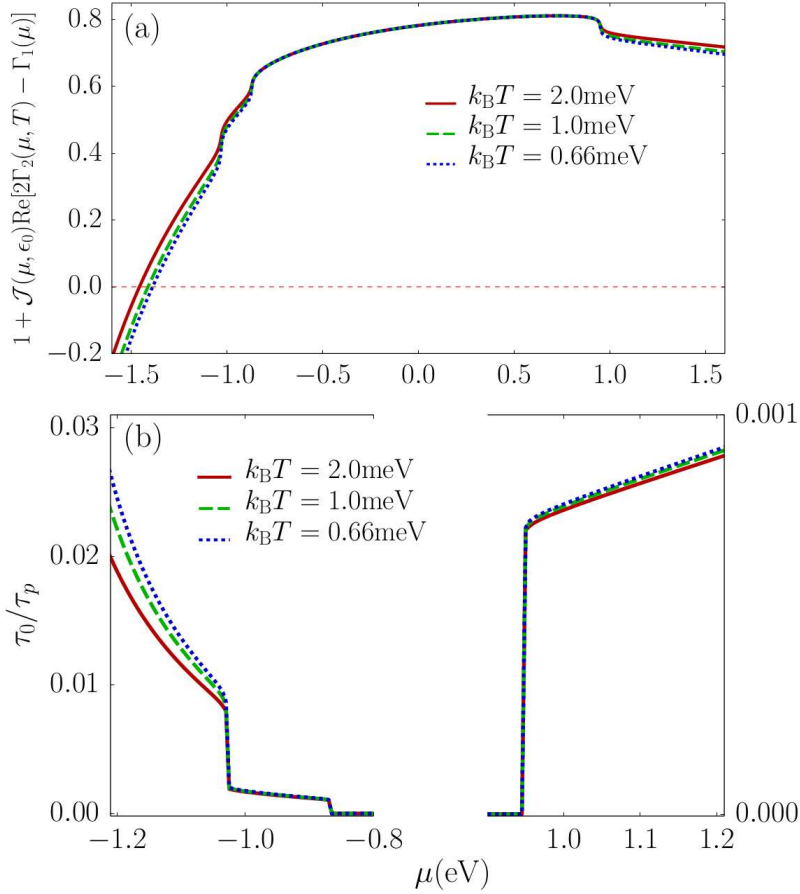}
\caption{(Color online) (a) Functional dependence of $1+{\cal J}(\mu,\epsilon_d){\rm Re}[2\Gamma_2(\mu,T_{K})-\Gamma_1(\mu)]=0$ as a function of the chemical potential, $\mu$. (b)  Momentum relaxation time as a function of the chemical potential. The effect of the spin-splitting of the valence band is obvious in both panels of this figure. Notice that
$\epsilon_d=-2.5$eV, $\tau_0=2\pi \hbar/( e{\rm V})$.}
\label{fig2}
\end{figure}

As a key parameter in determining of the Kondo temperature, the denominator in Eq. (\ref{parquet}) is plotted as a function of the chemical potential in Fig. \ref{fig2}a. Here, we assume an impurity level $\epsilon_d=-2.5$ eV deep inside the valence band and then the on-site Coulomb repulsion and the band width are considered to be $\sim 10$ eV. Inside the gap, since there is no thermally activated excitations, the function in the denominator does not depend on temperature whatsoever. It varies slightly with the chemical potential mostly originated from the $\mu$ dependence inside the coupling ${\cal J}$. In fact, the asymmetry with respect to $\mu=0$ (indicating the middle of the gap) can be understood simply from the fact that for $\mu>0$ we get further away from the impurity level and therefore the strength of coupling constant decreases, while for $\mu<0$ the Fermi level becomes closer to $\epsilon_d$ which leads to an increase in $\cal{J}$.
\par
When the Fermi level enters either the conduction or valence band, the screening owing to the exchange interaction with localized impurities switches on. In the highly doped cases, the resulting screening becomes stronger. This behavior is indicated by the decline in the denominator which equivalently leads to the shorter transport and spin lifetimes. As we see in Fig. \ref{fig2}(a), the slope of decrease in the valence band, especially for $\mu<-\Delta/2-\lambda$ is much higher which originates from the increase in the strength of the exchange coupling coefficient ($|{\cal J}|$) when the Fermi level is closer to the impurity level. More intriguingly, when the Fermi level is in the valence or conduction band due to the thermal excitations, the function varies with temperature. In particular for the particular set of parameters considered here and assuming typical few Kelvin temperatures ($k_BT\sim 1$), one sees that at some certain values of doping, the denominator vanishes revealing a divergence in the scattering rates as a signature of Kondo criticality. This is more clearly seen in Fig. \ref{fig2}(b) where the dimensionless scattering rate $\tau_0/\tau_p$
is plotted as a function of the chemical potential.
\par
As expected inside the gap the rate is vanishing and the lifetime is infinitely large while for doped cases the exchange coupling gives rise to finite scattering rate. An apparent strange observation is the fact that scattering rate is very small for the regimes Fermi level lies inside the spin split part of the valence band, while it is considerably larger in the conduction band and much stronger in valence when $\mu$ is below the lower spin subbands maxima. This unexpected behavior can be seen as a clear signature of the fact that in the spin split region in each valley only one type of spins does exist. Therefore, the exchange screening is significantly weak unlike the situations that carriers with both spins from both valleys are present at the Fermi level.
\subsection{Kondo temperature}
\begin{figure}[t]
\includegraphics[width=0.8\linewidth]{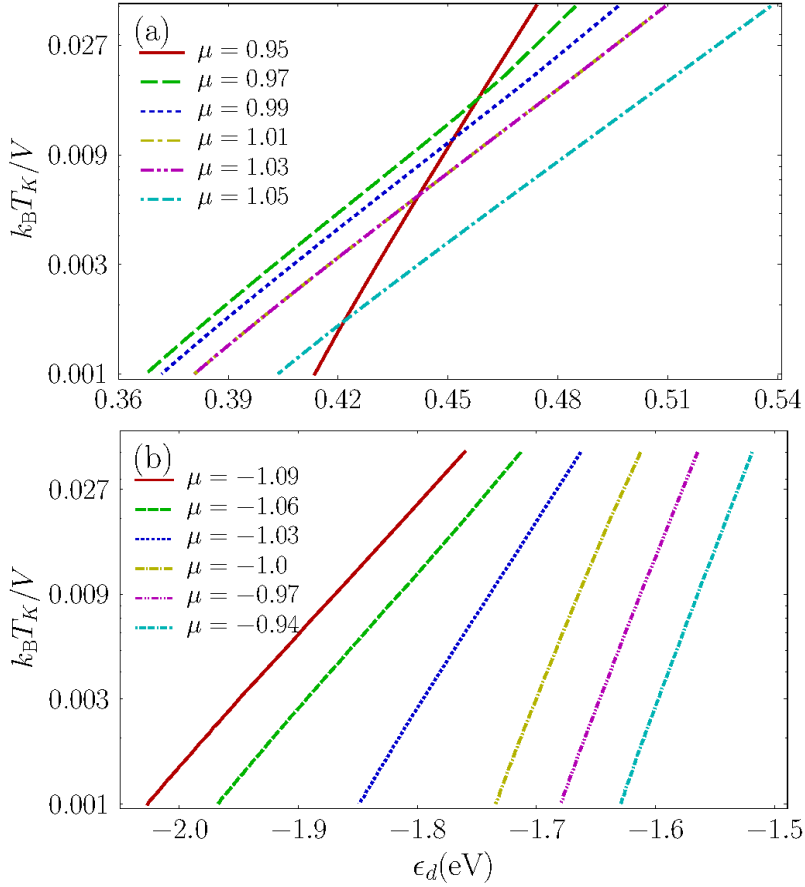}
\caption{(Color online) Kondo temperature of the monolayer MoS$_2$ as a function of
impurity level, $\epsilon_d$, for  the electron and hole doped cases are shown in the (a) and (b) panels, respectively. Note that the vertical axes have logarithmic scale.}
\label{fig3}
\end{figure}

In this part, the results for the dependence of the Kondo temperature on various parameters including chemical potential and impurity level are presented. In perturbative approach, $T_K$ is extracted numerically from the criterion of a divergent transport lifetime and subsequently vanishing denominator of Eq. (\ref{parquet}). The effect of the impurity level on the Kondo temperature for various values of the chemical potential $\mu$ is shown in Fig. \ref{fig3}.
Using  the logarithmic scale for $T_K$ the variations looks linear which is consistent with the expected exponential dependence of $T_K$ on the impurity level. In both electron- and hole-doped cases, the Kondo temperature increases by pushing the impurity level up to higher energies. This behavior mostly originates from increasing in the exchange coupling ${\cal J}_0$ when the impurity level becomes closer to the Fermi energy. Then a larger coupling constant leads to a higher Kondo temperature in accordance with the exponential dependence $T_K\propto \exp\{-1/|\rho(\mu){\cal J}_0|\}$. It must be mentioned that
the obtained results are justified when $\epsilon_{d}$ is well under $\mu$ and single occupancy is guaranteed. In fact, when the impurity level reaches to $\mu$, the coupling constant diverges and subsequently, the perturbation scheme breaks down. Therefore,  throughout the paper we assume only the case in which $\epsilon_d-\mu\gtrsim 1$ eV.
\begin{figure}[tp]
\includegraphics[width=0.8\linewidth]{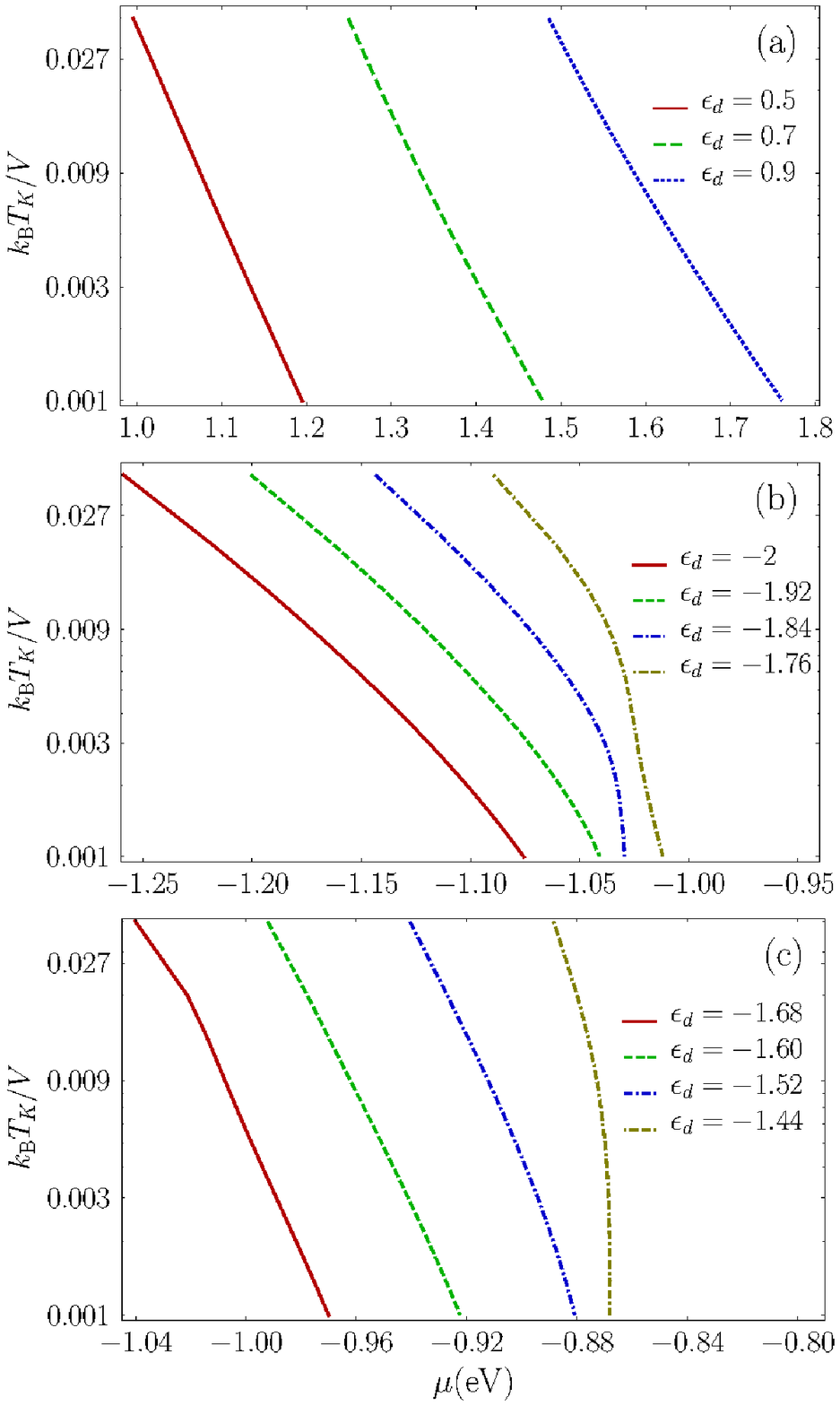}
\caption{(Color online) Kondo temperature of the monolayer MoS$_2$ as a function of
chemical potential, $\mu$, for  the electron doped case is shown in panel (a). Similar results for in the hole doped system are manifested in panels (b) and (c) corresponding the spin-degenerate and spin-split, respectively.  Note that the vertical axes denoting $T_K$ have logarithmic scale.}
\label{fig4}
\end{figure}
\par
In both electron- and hole-doped regimes, as one can see from Figs. \ref{fig3}(a) and \ref{fig3}(b), the slope of the increase in Kondo temperatures with $\epsilon_d$ is almost the same and slightly decreases with increasing the doping strength. The only exceptional case is $\mu=0.95$ with an unusually large slope of variations. In this case, the DOS is smaller since the Fermi level attains the conduction band minimum (CBM). Therefore $T_K$ drops faster by lowering the impurity level and subsequently decline in the coupling strength $|{\cal J}_0|$. But, one of the key findings of this work is the fact that even when the Fermi level lies inside the spin-split region $|\mu+\Delta/2|<\lambda$, the qualitative dependence of the Kondo temperature is the same as highly doped regimes. The only difference is in the slop of the dependence on $\epsilon_d$ which is slightly larger owing to the smaller value of the DOS in this region. This result can be interpreted as a result of the spin-flip scatterings accompanied by a change in the valley index. Therefore, despite of the spin-splitting, strong spin-flip scatterings between two valleys can occur.
\par
The variation of the Kondo temperature with the chemical potential ($\mu$) is depicted in Fig. \ref{fig4} using a logarithmic scale of $T_K$. The Kondo temperature always increases monotonically by lowering the chemical potential. In particular, we see almost the linear dependence of $T_K$ in the logarithm scale on the chemical potential, except for the doping which is very close to the two valence subband edges ($\mu=-\Delta/2\pm\lambda $). This behavior is followed from the general exponential dependence of the Kondo temperature on the inverse of the coupling constant multiplied by the DOS. Either by increasing the impurity level or by decreasing the doping, which subsequently leads to a larger $|{\cal J}_0|$ value, the Kondo effect becomes more profound indicated by a higher value of the Kondo temperature. Deviations from the linear dependence of $\log T_K$ on $\mu$ originates from the DOS variation with the chemical potential which is more drastic close to the band edges as shown in Figs. \ref{fig5}(b) and \ref{fig5}(c). Although the coupling constant always decreases by pushing the Fermi level to the higher energies, the DOS increases (decreases) with $\mu$ in the conduction (valence) band. Therefore, it can be understood why the Kondo temperature increases faster when the Fermi level is in the valence rather than the conduction band, according to Figs. \ref{fig4}(a) and \ref{fig4}(c).
\par
So far, we have discussed over the onset of the Kondo effect using the perturbation theory for temperatures above $T_K$. Then the Kondo temperature is obtained from the breakdown point of perturbation procedures in which all the transport signatures and particularly the exchange induced scattering rate diverges. This method has the advantage that the full low-energy band structure is completely taken into account. In the remainder of this part
we give the results of the poor man's scaling method for the Kondo temperature which as an alternative method can justify the results of the perturbation theory. As it has been explained at the end of Sec. \ref{sec2}, this analysis is implemented using a slightly simplified model for the band structure of TMDs-ML.
\par
Using the RG flow Eq. (\ref{rg-flow}) the Kondo temperatures for three different doping regimes are obtained as below,
\begin{eqnarray}
k_BT_k&=& e^{ \frac{-1}{2\rho_c^0 |\mu {\cal J}_0|}\frac{D_0+\mu}{4\mu-\Delta}},~~~~~~\mu>\Delta/2 \nonumber\\
k_BT_k&=& e^{ \frac{-1+d_\lambda}{\rho_v^0 |\mu {\cal J}_0|}\frac{D_0+|\mu|}{4|\mu|-\Delta+2\lambda}},~~|\mu+\frac{\Delta}{2}|
<\lambda\\
k_BT_k&=& e^{  \frac{-1}{2\rho_v^0 |\mu {\cal J}_0|}\frac{D_0+|\mu|}{4|\mu|-\Delta+2\lambda} }
,~~\mu<-\Delta/2-\lambda \nonumber
\end{eqnarray}
with
$d_\lambda=2\lambda/\left(4|\mu|-\Delta/2+\lambda\right)$ which is small ($d_\lambda<0.06$) for typical values of doping $|\mu|\sim 1$eV.
The original bandwidth and bare coupling constant are indicated by $D_0$ and ${\cal J}_0$, respectively. More details about these calculations can be found in Appendix \ref{app-c}. We see that the behavior of the Kondo temperature with doping is very similar in the \emph{n} and heavily \emph{p}-doped cases. In addition, the Kondo temperature in the spin-split region differs only by a factor of one half in the exponent which originates from the fact that only one spin subbands contributes effectively in the exchange induced scatterings. Putting all these results together, the qualitative agreement between the results obtained from poor man's scaling analysis and the perturbation theory is well justified.

\subsection{Kondo conductivity}
In this part we study the magnetic impurities contribution in the conductivity in the vicinity of the Kondo regime which is called the Kondo conductivity. The conductivity can be calculated from the Drude relation and for the contribution we are interested in the transport lifetime $\tau_p$ is given by (\ref{parquet}) which takes the effect of exchange interaction (\ref{exchange}) into account,
\begin{equation}
\sigma^{\rm ex}_{K} =  e^2 \sum_{\tau , s}\int^{D}_{-D} d\epsilon  \rho_{\tau s}(\epsilon)  v^2_{\tau s}(\epsilon) \tau_p (\epsilon)
 \left [ -\frac{\partial f(\epsilon)}{\partial \epsilon} \right ],
\end{equation}
where $v_{\tau s}(\epsilon)=(1/\hbar)\partial \epsilon_{k\tau s}/\partial k_x |_{\epsilon_{k\tau s}=\epsilon}$ and $\rho_{\tau s}(\epsilon)$ is the density of states for a given spin ($s$) and valley ($\tau$) index.
\par
\begin{figure}[tp]
\includegraphics[width=0.8\linewidth]{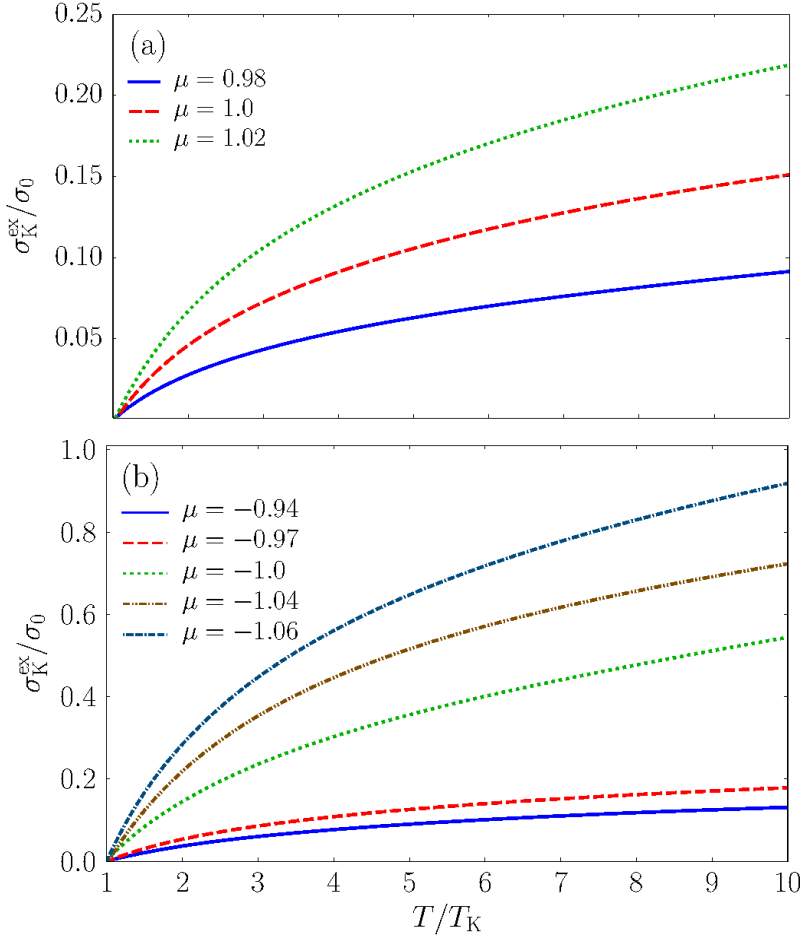}
\caption{(Color online) Temperature dependence of Kondo conductivity originating from resonant scattering by magnetic impurities in (a) conduction and (b) valence band. The impurity level for different cases in which the chemical potential lies inside the conduction band, spin split region of valence band, and the region where both spin subbands are present (below the spin split region) is $\epsilon_d=0.4, -1.65, -1.85$ eV, respectively.}
\label{fig5}
\end{figure}

The variation of the Kondo conductivity with temperature, which is scaled with respect to $\sigma_0=(e^2/\hbar)[32 \pi/(3n_{\rm imp} \Omega)]$, is shown in Fig. \ref{fig5} for various values of doping. In general, the Kondo conductivity decreases as temperature decreases and it vanishes upon reaching the Kondo temperature. Such behavior originates from the fact that at low temperatures, the strong scattering from the impurities caused by the exchange interaction leads to the shorter transport lifetime $\tau(\mu)$ and subsequently the magnetic impurities contribution in resistivity increases drastically in the vicinity of the $T_K$. It must be noticed that very close to the Kondo temperature and below it, our perturbative analysis based on the parquet approximation breaks down. However, the qualitative behavior above the Kondo temperature can be justified.
When the temperature is scaled with $T_K$ the Kondo conductivity is always an increasing function of doping strength. Therefore, at fixed temperature with respect to $T_K$, the conductivity $\sigma_K^{\rm ex}$ increases by increasing doping strength. This picture is completely consistent with the Boltzmann transport theory where the conductivity increases by increasing the carrier density. It must be noticed that the effect of the chemical potential variation in the transport life time is mainly absorbed inside the Kondo temperature. Therefore by scaling the temperature with $T_K$, only the effect of carrier density appears in the conductivity.
\par
It must be noticed that although $\sigma_{K}^{\ ex}$ indicates the main features of Kondo screening, one should consider the effect of scattering by ordinary (nonmagnetic) impurities and also electron-phonon interactions in the total conductivity. These two extra contributions dominate at zero temperature and very high temperatures, respectively, and are not considered here. From our result in the vicinity of a strong Kondo screening regime, which the scattering due to the exchange interaction with magnetic impurities is dominant, the total resistivity increases by lowering the temperature. On the other hand, it is known that at higher temperatures ($T\gg T_K$) the electron-phonon contribution can lead to a $T^4$ dependence of the resistivity over temperature \cite{hewson}. Then from the combination of these two contributions in the resistivity, one can easily expect that the appearance of the minimum resistivity at a temperature $T_{\rm min}$ which is usually larger than $T_K$. It should be noticed that in contrast to the Kondo temperature, $T_{\rm min}$ can be varied with the concentration of impurities and also the parameters related to the electron-phonon iterations.

\subsection{Comparison with related studies}
It is worth commenting on the relation between our results with previous investigations about the Kondo effect in 2D materials, including two dimensional electron gas (2DEG), graphene, and topological surface states. First of all, in the case of 2DEG due to the constancy of DOS, the Kondo temperature is not tunable via doping whatsoever. However, recently an enhancement of the Kondo temperature with Rashba SOI has been predicted at constant low fillings, due to the emergence of Dzyaloshinsky-Moriya type exchange term \cite{zarea12,sandler16}. In contrast to conventional 2DEG, theoretical investigations have suggested strong doping dependence of the Kondo effect in graphene \cite{vojta}. Remaining under some debates, there are also experimental evidences for the tunability of $T_K$ with chemical potential~\cite{fuhrer,fuhrer-cmt}. On the other hand, it has been shown that Kondo effect in graphene can show a quantum critical behavior with power low dependence of $T_K$ on the coupling constant \cite{uchoa11}. Such behavior mostly originates from the linear dispersion as well as the existence of Dirac points in graphene \cite{vignale15}. Theoretical and experimental investigations about Kondo features in surface states of topological insulators which are doped with magnetic impurities has been carried out by various groups \cite{vojta13,isaev15}. In particular, it has been revealed that despite the spin-momentum locking, the magnetic impurities are Kondo screened by the surface-state electrons which have Dirac dispersion \cite{vojta13}.
\par
In contrast to all above mentioned gapless 2D systems, TMDs-ML have a dispersion relation similar to gapped Dirac systems besides the spin splitting present in the valence band with different sign for the different valleys. Then, as it has been discussed before, the transport properties and Kondo physics at the presence of magnetic impurities are mostly governed by gapped quadratic spectrum of the TMDs-ML and particularly the linear energy dependence of DOS. Unlike graphene and other gapless systems no quantum critical behavior is observed and irrespective of the strength of ${\cal J}_0$ the system is always in the strong coupling limit revealed by the RG equations obtained from poor man's scaling. On the other hand the Kondo temperature is weakly influenced by the spin splitting. In particular in the spin split regime of doping the Kondo effect is slightly weakened due to the decline in DOS. The origin of behavior is owing to the presence of valley degree of freedom and intervalley scatterings providing the channel for spin-flip processes and subsequently the Kondo screening.
\par
Finally, regarding the possible experimental investigations, we note that in the context of graphene, the localized magnetic moments can be usually formed via the defects \cite{fuhrer,uchoa11,vojta}. Hence, the Kondo physics of graphene despite the various studies so far, needs to be understood more clearly by further experimental and theoretical studies. Nevertheless, in TMDs the magnetic adatoms as well as defects can lead to the localized magnetic moments which has been verified by {\it ab initio} studies \cite{cheng13,vaveh13,yakobson,idrobo, yan14,Sachs13}. In addition, the ionic liquid gating method provides an extreme tunability of the chemical potential in TMDs-ML \cite{morpurgo}. Therefore we can conclude that it is quite feasible to experimentally check the results obtained here for the Kondo effect and minimum resistivity.
\section{Conclusion}\label{sec4}
In summary, we have used perturbation theory to investigate the effect of dilute magnetic impurities in the Kondo characteristics and spin relaxation inside a monolayer of transition metal dichalcogenides. It is shown that inside the valence band, due to the large spin splitting, the transport lifetime and spin relaxation are much longer when the Fermi level is in the spin split region in comparison with situations in which the chemical potential is deep inside the valence band and crosses both spin subbands. In strong contrast, although the lowest orders of spin-dependent scattering is strong only when the Fermi level lies deep in the conduction or valence band, however, the Kondo effect is not suppressed even in the spin-split region. Such behavior, which is also checked using poor man's scaling analysis, originates from the many body character of the Kondo effect and the fact that in the Kondo screening, electrons can be equally scattered inside each valley or between them. We further investigate the effect of doping and impurity levels on the Kondo temperature as well as the temperature dependence of the magnetic impurities contribution in the conductivity. Although it is not explicitly considered here, combining the contribution of magnetic impurities with that of electron-phonon scatterings which is dominant at higher temperatures, one would expect the appearance of a minimum at the resistivity as a function of temperature.
Finally the possible experimental verification of the Kondo effect in transition metal dichalcogenides is discussed.
\ack
This work is partially supported by Iran Science Elites Federation under Grant No. 11/66332.
H.R. acknowledge funding from the European Commission under the Graphene Flagship, contract CNECTICT-604391. and partial support by MIUR (Italy) through the program ``Progetti Premiali 2012'' - Project ``ABNANOTECH".
\appendix
\section{Kondo versus Anderson Model}\label{app-a}
In order to obtain Kondo Hamiltonian we use the Schrieffer-Wolff transformation as $\tilde{{\cal H}}=e^S{\cal H}e^{-S}$ assuming ${\cal H}_V$ as perturbation.
Then the transformed Hamiltonian takes the form,
\begin{eqnarray}
\tilde{{\cal H}}={\cal H}_{\rm bath}+{\cal H}_{\rm imp}+\frac{1}{2}[S,{\cal H}_V]+\cdots
\end{eqnarray}
provided by the constraint $[{\cal H}_{\rm bath}+{\cal H}_{\rm imp},S]={\cal H}_V$
to fulfill the second order perturbation in hybridization $V_{k\tau d}$.
Using the suitable transformation generator $S$, the second order terms of the Hamiltonian read,
\begin{eqnarray}
&&{\cal H}_2=\frac{1}{2}[S,{\cal H}_V]={\cal H}'_{\rm imp}+{\cal H}_{ex}+{\cal H}_{dir}+{\cal H}_{ch}
\\
&&{\cal H}'_{\rm imp}=-\sum_{k\tau s}{[W^{\tau\tau}_{kks}+\frac{1}{2} J^{\tau\tau}_{kks} n_{d\bar s}]n_{ds}}
\\
&&{\cal H}_{\rm ex}=-\frac{1}{2}\sum_{kk'\tau \tau's}
[J^{\tau'\tau}_{k'ks}c^\dagger_{k'\tau' s}c_{k\tau\bar s}c^\dagger_{d\bar s}c_{ds}
 \nonumber\\
&&~~~~~~~~~+I^{\tau'\tau}_{k'ks}\frac{n_{ds}-n_{d\bar s}}{2}c^\dagger_{k'\tau' s}c_{k\tau s}] \\
&&{\cal H}_{\rm dir}=\sum_{kk'\tau \tau's}{[W^{\tau'\tau}_{k'ks}+I^{\tau'\tau}_{k'ks}
\frac{n_{ds}+n_{d\bar s}}{4}]c^\dagger_{k'\tau' s} c_{k\tau s}}
\\
&&{\cal H}_{\rm ch}=-\frac{1}{4}\sum_{kk'\tau \tau's}{{\tilde{J}}^{\tau'\tau}_{k'ks}c^\dagger_{k'\bar s}c^\dagger_{ks}c_{ds}c_{d\bar s}}+H.C.
\end{eqnarray}
with Kondo couplings,
\begin{eqnarray}
&&J^{\tau'\tau}_{k'ks}=V_{k'\tau'd}V^\ast_{k\tau d}[\frac{1}{\epsilon_{k\tau\bar s}-\epsilon_{+}}+\frac{1}{\epsilon_{k'\tau' s}-\epsilon_{+}}
\nonumber\\
&&~~~~~~~  -\frac{1}{\epsilon_{k\tau\bar s}-\epsilon_{-}}-\frac{1}{\epsilon_{k'\tau' s}-\epsilon_{-}}]\\
&&I^{\tau'\tau}_{k'ks}=V_{k'\tau'd}V^\ast_{k\tau d}[\frac{1}{\epsilon_{k\tau s}-\epsilon_{+}}+\frac{1}{\epsilon_{k'\tau' s}-\epsilon_{+}}
\nonumber\\
&&~~~~~~~  -\frac{1}{\epsilon_{k\tau s}-\epsilon_{-}}-\frac{1}{\epsilon_{k'\tau' s}-\epsilon_{-}}] \\
&&W^{\tau'\tau}_{k'k s}=\frac{1}{2}V_{k'\tau'd}V^\ast_{k\tau d}[\frac{1}{\epsilon_{k\tau s}-\epsilon_{-}}+\frac{1}{\epsilon_{k'\tau' s}-\epsilon_{-}}]
\end{eqnarray}
with the property $V_{k\tau d}\tilde{J}^{\tau'\tau}_{k'k s}=V^\ast_{k\tau d}J^{\tau'\tau}_{k'k s}$ and definitions $\epsilon_{+}=\epsilon_d+U$, $\epsilon_{-}=\epsilon_d$.
\section{The second order $T$-matrix terms}\label{app-b}
All of the second order $T$-matrix terms versus $I$ and $J$, will contain a summation over the intermediate (virtual) states $k',\tau'$. Moreover depending on the character of the virtual states as an electron or a hole, their corresponding occupancies described by the Fermi distributions will appear.
Replacing the summation over the momenta with integration over the energy and using the simplified notation as $\epsilon_\sigma:=\epsilon_{k\tau \sigma}$ and $\epsilon'_\sigma:=\epsilon_{k'\tau'\sigma}$, the following expressions will be obtained,
\begin{eqnarray}
T^{(2)}_{ss} &=& \frac{\Omega|V|^4}{4} [2 s I_2(\epsilon_{s},\epsilon'_{s}) S_z+I_1(\epsilon_{s},\epsilon'_{s}) (S^2-s S_z)],\nonumber\\
T^{(2)}_{\bar ss} &=& \frac{\Omega|V|^4}{4}[2I_2(\epsilon_{s},\epsilon'_{\bar s})-I_1(\epsilon_{s},\epsilon'_{\bar s})]S_{s},
 \end{eqnarray}
in which,
\begin{eqnarray}
I_1(x,y) &=& \int^{D_0}_{-D_0} dz \rho (z)  Q(x,y,z),\nonumber\\
I_2(x,y) &=& \int^{D_0}_{-D_0} dz \rho (z)  Q(x,y,z) f(z),\label{integrals}
 \end{eqnarray}
with $D_0$ denoting the band cutoff for the Kondo coupling.
The function $Q(x,y,z)$ is a rational function with respect to all three variables and can be decomposed as,
\begin{equation}\label{eq:Q}
Q(x,y,z)=\frac{Q_{0}(x,y)}{x-z+i\eta}+ \sum_{l=+,-} \left[   \frac{Q^{l}_1(x,y)}{z-\epsilon_{l}}+
 \frac{Q^{l}_2(x)}{(z-\epsilon_{l})^2}\right],
\end{equation}
in which the other functions are
\begin{eqnarray}
&&Q^l_2(x)=\frac{1}{x-\epsilon_l},\nonumber\\
&&Q^l_1(x,y)=\left[ R(x) +R(y)-\frac{2}{U} \right]Q^l_2(x) +l \left[Q^l_2(x) \right]^2,\nonumber\\
&&Q_0(x,y)=2R(x)\left[R(x)+ R(y)\right],
\end{eqnarray}
with $R(x)=Q^{+}_2(x)-Q^{-}_2(x)$.
As it has been mentioned before the perturbative approach results in the rational dependence with respect to the unperturbed energy differences. Then in order to keep the consistency of perturbation the impurity level $\epsilon_d$ and host electron energy, namely the chemical potential $\mu$ should be far from each other. Therefore, without any problem we can substitute $\epsilon_\sigma-\epsilon_l+i\eta$ with $\epsilon_\sigma-\epsilon_l$. One can immediately see that the last two terms in (\ref{eq:Q}) diverge when the host electrons energy hits the impurity level. But this is of course beyond the validity of the perturbative formalism. On this ground and for the sake of consistency and simplicity, we do not take these terms into account. Moreover, since the initial and final state energies are equal we immediately see the relation $Q_0(x,x)=[2R(x)]^2$. This assumption leads to an effective coupling coefficient ${\cal J}$ instead of all coupling coefficients $I^{\tau'\tau}_{k'k s}$ and $J^{\tau'\tau}_{k'ks}$ in the Hamiltonian (\ref{exchange}) and is given by,
\begin{equation}
{\cal J}(\epsilon,\epsilon_d)=|V|^2\sqrt{Q_0(\epsilon,\epsilon_d)},
\end{equation}
which finally yields to the expression (\ref{J}) in the main text.

%%%%%%%%%%%%%%%%%%%%%%%%%%%%%%%%%%%%%%%%%%%%%%%%%%%%%%%%%%%%%%%%%%%%%%%%%%%%%%
%%%%%%%%%%%%%%%%%%%%%%%%%%%%%%%%%%%%%%%%%%%%%%%%%%%%%%%%%%%%%%%%%%%%%%%%%%%%%%
%%%%%%%%%%%%%%%%%%%%%%%%%%%%%%%%%%%%%%%%%%%%%%%%%%%%%%%%%%%%%%%%%%%%%%%%%%%%%%


\begin{thebibliography}{2}
\bibitem{footnote}
In order to obtain the constants we have used the set of band parameter given in Ref. 57 which provides  a good fit with the empirical or ab initio band structure close to the K points.
\end{thebibliography}
\end{document}